\documentclass[runningheads]{llncs}
\usepackage[T1]{fontenc}
\usepackage{graphicx}
\usepackage{verbatim}
\usepackage{tabularx}
\usepackage{multirow}
\usepackage{subcaption}
\usepackage{caption}
\usepackage{float}
\usepackage{wrapfig}
\usepackage{booktabs}
\usepackage{nomencl}
\makenomenclature
\usepackage{minted}
\usepackage{cleveref}
\usepackage{comment}

\begin{document}
\title{Beyond Problem Solving: Large Language Models for Emotional and Reflective Support in Mathematics Learning}
\titlerunning{Beyond Math Problem Solving: LLMs for Emotional and Reflective Support}
%
\author{Vera Rief\inst{1}\orcidID{0009-0001-5787-3748} \and
Mirella Hladký\inst{1}\orcidID{0000-0001-5565-3998} \and
Minju Yoo\inst{2}\orcidID{0009-0000-2964-0464} \and
Stephanie Heel\inst{1}\orcidID{0009-0005-3297-7590} \and
Shintaro Sato\inst{1}\orcidID{0009-0002-8790-9411} \and
Tomohiro Nagashima\inst{1}\orcidID{0000-0003-2489-5016}}
\authorrunning{V.Rief et al.}
%
\institute{
Saarland University, Saarland Informatics Campus, Saarbrücken, Germany
\email{veri00001@uni-saarland.de, mirella.hladky@uni-saarland.de, heel.stephanie@gmail.com, sato@cs.uni-saarland.de, nagashima@cs.uni-saarland.de} \and
Korea Advanced Institute of Science and Technology, Daejeon, South Korea
\email{minjuyoo@kaist.ac.kr}
}

\maketitle              
\begin{abstract}
Intelligent Tutoring Systems (ITSs) traditionally focus their adaptive support on cognitive aspects of learning. Although effective, little is known about how such systems can be enhanced by addressing students’ emotional states. In particular, the role of mindful interventions for supporting student learning and experiences in adaptive math learning remains underexplored. We developed "Math with Matt", an ITS that leverages Large Language Models (LLMs) to provide both cognitive and emotional support in algebra learning. The system offers 1) an LLM-based mindful chat that delivers context-sensitive emotional support through a pedagogical agent Matt, and 2) mindful feedback and hint messages (not just evaluative) to enhance learning experiences and reduce math anxiety. We conducted a classroom study with 7th graders, comparing a Mindful version against a version with cognitive support only. Overall, the ITS reduced executive state-math anxiety and improved students' math learning, though no significant differences emerged between the conditions. However, students with the mindfulness interventions showed higher learning efficiency and well-balanced problem-solving behavior, since they achieve a similar level of math learning with less learning time and fewer requested hints compared to the Cognitive version. Additionally, they reported that the pedagogical agent felt more supportive and caring than students in the cognitive condition. Our study demonstrates the feasibility and scalability of integrating mindfulness into ITSs through LLM-based interactions and positions LLMs as an adaptive, socio-emotional layer within cognitive math tutoring. 

\end{abstract}
%
%
%

\section{Introduction}
Math anxiety (MA) is an unpleasant emotional state characterized by fear and tension that emerges in situations involving numerical processing or mathematical problem solving \cite{Richardson1972_MARS}. It is a widespread phenomenon \cite{OECD2015} that negatively affects students by impairing working memory, lowering performance, and reducing engagement with mathematical tasks \cite{Ashcraft2001_Anx_and_Perform}. In particular, state-MA, which occurs during math-related situations, has been reported to have a negative correlation with math performance \cite{Orbach2019_StateTrait}. 

Existing interventions addressing this negative relationship between anxiety and math performance are typically not integrated into regular instruction, limiting their scalability. For instance, mindfulness-based interventions, such as guided breathing exercises at the beginning of class, provide emotion-regulation support and have been shown to effectively reduce MA in real time \cite{Zenner2014,Zuo2023}. However, these interventions typically rely on instructor-led (with training), multi-session programs that are challenging to embed into digital/non-digital learning environments \cite{Zenner2014}. Also, in the context of technology-enhanced learning, Intelligent Tutoring Systems (ITSs) and pedagogical agents offer strong cognitive and motivational support, yet they rarely attend to students' emotional states during problem solving \cite{Koedinger2007,Schroeder2025_Meta,Lehn2011}. Recent advances in Large Language Models (LLMs) enable flexible and context-sensitive conversational support, but their integration into structured tutoring environments still remains limited \cite{Wollny2021_chatbotreview}.

To address these gaps, we developed an ITS that leverages LLMs in providing both cognitive and mindfulness scaffolding in algebra learning to support students' math learning while reducing anxiety. 
Our system offers 1) an LLM-based mindful chat that delivers context-sensitive emotional support through a pedagogical agent, as well as guided breathing exercises, and 2) cognitive-mindful feedback and hint messages aimed to enhance learning processes, learning experiences, while alleviating MA. 
We conducted a classroom study comparing this ITS with a version that provides only cognitive support to test its effects on students' MA, math learning, and perceived usability and learning experiences.

This study demonstrates that an ITS can be enriched with LLM‑driven mindfulness interventions, while still providing the cognitive hints and feedback that support math learning. It showed the feasibility and scalability of such interventions and that they can be delivered automatically, in real time, and without the need for specialized instructor training.
 
\section{Background and Related Work}
\subsection{Math Anxiety}
Math anxiety (MA) is an emotional reaction with feelings of worry, stress, and powerlessness when faced with mathematics lessons, homework, or tests \cite{Hembree1990_Meta,OECD2015}, which affects approximately one third of students worldwide \cite{OECD2015}. MA is often treated as two separate yet closely-related sub-constructs: trait-MA (a stable disposition) and state-MA (a context-dependent reaction) \cite{Orbach2019_StateTrait}. State-MA can be further differentiated into\textit{ anticipatory MA}, which occurs when students are expecting to participate in math tasks, and \textit{executive MA}, which occurs while they are actively working on math tasks \cite{OECD2015}.
MA is closely associated with perceived lower math performance relative to peers \cite{OECD2015,Hembree1990_Meta,Zuo2023}. It negatively affects performance by consuming cognitive resources during task execution, reducing working-memory capacity \cite{Ashcraft2001_Anx_and_Perform}. Research consistently demonstrates a negative relationship between MA, especially state-MA \cite{Orbach2019_StateTrait}, and math performance \cite{Hembree1990_Meta,OECD2015,Richardson1972_MARS}. 
To support math-anxious students, prior research has demonstrated that cognitive and emotion-focused interventions that target anxiety during task execution, especially short-relaxation-based practices, have shown promise in reducing MA and restoring performance in classroom settings \cite{Brunye2013,Hembree1990_Meta,Sammallahti2023_Meta}.

\subsection{Mindfulness Interventions}
Mindfulness is generally understood as the continuous, non-judgmental awareness of the present moment that develops through deliberate, non-reactive, openhearted attention \cite{Kabat-Zinn2015}. 
Research has supported the efficacy of mindfulness-based intervention (e.g., breathing and meditation practices) in treating various psychological problems and disorders, such as perceived stress or depression \cite{Baer2009_Rumination}. Their effects are attributed to the de-automatization of habitual cognitive and emotional processes \cite{Kang2012_De-Auto}, which enhances cognitive control and supports adaptive self-regulation. Additionally, mindfulness reduces rumination and emotional avoidance, and supports constructive behavior even in the presence of unpleasant thoughts or emotions \cite{Baer2009_Rumination}. 
Kabat-Zinn’s seven principles of mindfulness (\textit{Non-Judgment, Patience, Beginner’s Mind, Trust, Non-Striving, Acceptance, and Letting Go}) \cite{Kabat-Zinn2015} offer a framework to implement OECD's \cite{OECD2024} suggestions to prevent MA through a non-judgmental, curious, approach to math, that values process over outcome, allows an individual pace, and trust their own abilities. 

For classroom application, brief, focused breathing practices are feasible and have been reported to have an association with improvements in cognitive performance \cite{Zenner2014}. Recent studies suggest that guided breathing before a math task (targeting state-MA) can reduce negative emotions, increase calmness and improve performance \cite{Brunye2013,Hladky2025_MindfulMathTutor,Zuo2023}. 
However, these existing solutions (e.g., guided breathing) often require trained professionals and multiple sessions, which do not scale effectively.
In this context, integrating mindfulness into digital learning environments represent a potentially cost-effective and scalable approach for supporting students' well-being and improving learning outcomes \cite{Zenner2014}. However, this area remains relatively unexplored as most systems lack explicit socio-emotional support \cite{Bereczki2024}.

\subsection{Intelligent Tutoring Systems and Pedagogical Agents}
Intelligent Tutoring Systems (ITSs) are interactive, computer-based learning environment that use artificial intelligence (AI) techniques to provide personalized instruction and adaptive feedback to students \cite{Gobert2022_ITS,Koedinger2007,Lehn2011}. Research indicates that ITS-supported learning can effectively enhance students' performance, improve traditional classroom teaching, and is often more effective than teacher-led instruction, large group formats and non-adaptive digital learning tools \cite{Gobert2022_ITS,Koedinger2007}. 
By integrating pedagogical agents, which are virtual characters that guide and engage with learners, providing a sense of social presence \cite{Clarebout2012,Schroeder2025_Meta}, ITSs can support learners not only cognitively but also socio-emotionally, thus enhancing learning outcomes and motivation~\cite{Schroeder2025_Meta}. 
A recent study by Hladký et al. has found that a hard-coded digital learning environment incorporating a pedagogical agent designed around mindfulness effectively reduced students’ MA~\cite{Hladky2025_MindfulMathTutor}.

These benefits could be further enhanced by leveraging the capability of Large Language Models (LLMs) to enable natural, flexible, and adaptive interactions with pedagogical agents~\cite{Stamper2024_llmbasedfeedback}. 
Kumar et al. demonstrated that brief supportive conversations with a GPT-3-based chatbot can improving users' mood and mental well-being \cite{Kumar2022_prompt}. 
Existing ITSs excel at cognitive scaffolding yet rarely embed real‑time socio‑emotional support, such as mindfulness, leaving state-MA largely unaddressed. Leveraging LLM-based pedagogical agents to deliver adaptive support within ITS potentially offers a scalable classroom solution that simultaneously increases learning and reduces MA. 
\section{Design of the Mindful Intelligent Tutor for Algebra}
To address the lack of scalable mindfulness interventions for supporting math anxiety and learning, we developed an LLM-enhanced ITS that provides cognitive, domain-specific support on linear-equation systems with mindfulness interventions supporting emotional regulation. The ITS was implemented using the Cognitive Tutor Authoring Tools (CTAT) \footnote{\url{https://github.com/CMUCTAT/CTAT}} \cite{Aleven2006_CTAT}. 
The content of the ITS is divided into discrete chapters (introduction, three math learning units, and closing) and allows students to proceed at their own pace. 
To provide a learning experience where students could receive both cognitive and mindful support, we designed a fox-shaped pedagogical agent called "Matt" who uses human-like expressions and converses in first- and second-person language ("you," "we") (see the following section to learn more about Matt).

\subsection{Math Learning Units}
\begin{figure}[htpb]
    \centering
     \begin{subfigure}[t]{0.485\textwidth}
        \includegraphics[width=\linewidth]{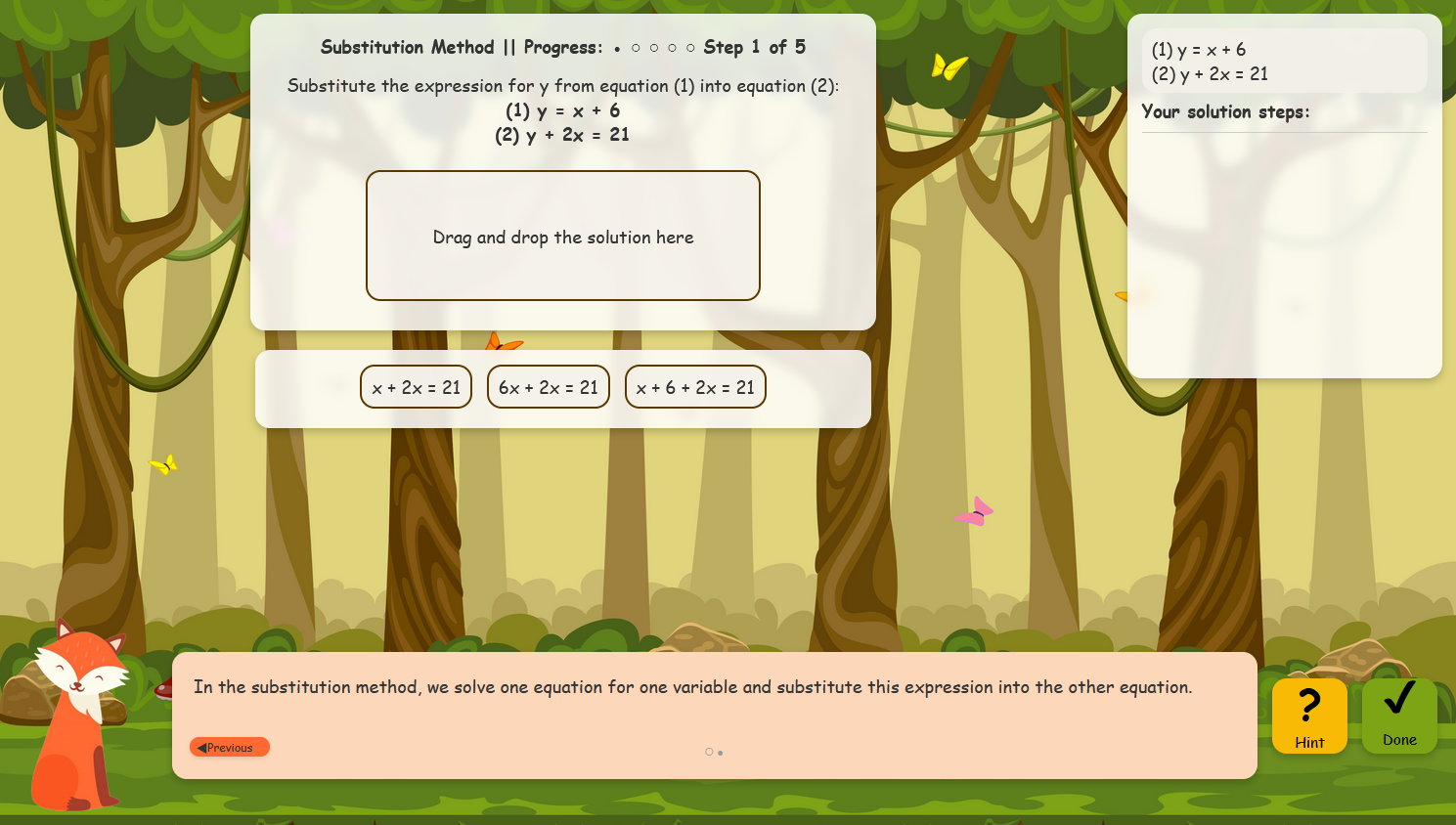}
        \caption{Drag and drop interaction with three options and Matt giving instructions}
        \label{fig:drag}
      \end{subfigure}
      \hfill
      \begin{subfigure}[t]{0.475\textwidth}       
        \includegraphics[width=\linewidth]{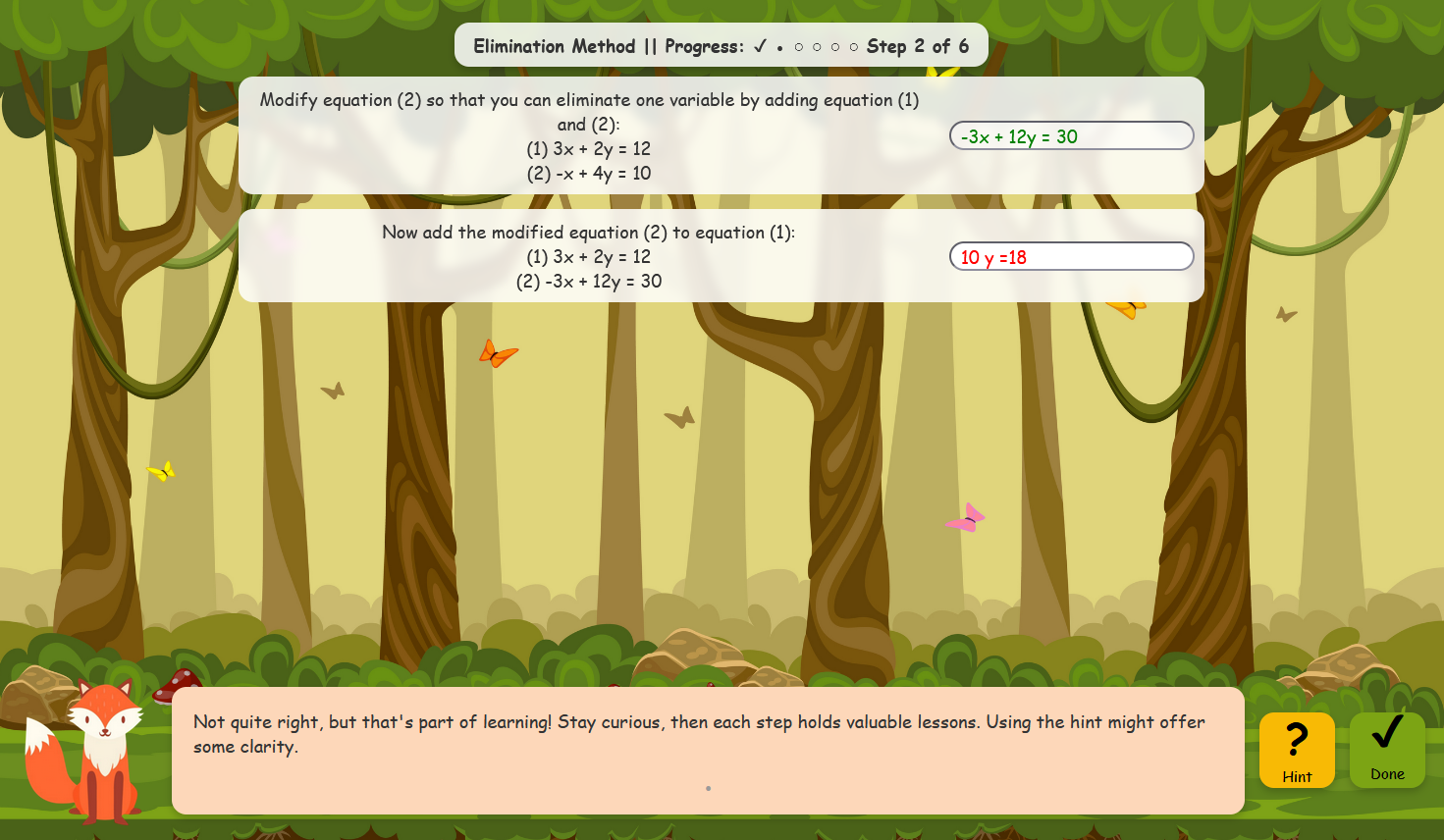}
        \caption{Free-text input interaction with Matt giving error feedback using mindful language}
        \label{fig:textinput}
      \end{subfigure}
  \caption{Math interfaces}
  \label{fig:chats}
\end{figure}

The ITS includes three math units for learning to use different solution methods for linear systems of equations. Drag-and-drop interaction is used in Unit 1 where students can practice the substitution method (one equation is first solved for a variable and substituted into the other equation) while familiarizing themselves with the tutor. In Unit 1, students choose from three options showing next steps to answer each step and drag an option to the field above to evaluate the answer (Fig. \ref{fig:drag}). This interaction provides scaffolded interactions where students do not have to type in answers. Units 2 involves the elimination method (where linear combinations of the equations are used to cancel one variable first) and Unit 3 involves the equalization method (the same variable is isolated in both equations first and the resulting expressions are equated). Both Units 2 and 3 require students to type in equation-solving steps by themselves (Fig. \ref{fig:textinput}). 
Every unit contains three problems, resulting in nine problems (with a total of 57 problem-solving steps) plus ten optional bonus exercises, where the mixture of the three solution methods are practiced (additional 65 steps). Every unit contains three problems, resulting in nine problems (with a total of 57 problem-solving steps) plus ten optional bonus exercises, where the mixture of the three solution methods are practiced (additional 65 steps).
Each problem-solving step offers three levels of hints, and students would receive immediate success or failure feedback by Matt after each attempt. Additionally, students' problem-solving history remains visible on the screen. Visual cues, such as bold fonts and highlighting, are used throughout the tutor to guide student's attention during learning.

\subsection{Mindfulness Interventions: Dynamic Chats, Breathing and Mindful Language}
\begin{figure}[htpb]
    \centering
     \begin{subfigure}[t]{0.47\textwidth}
        \includegraphics[width=\linewidth]{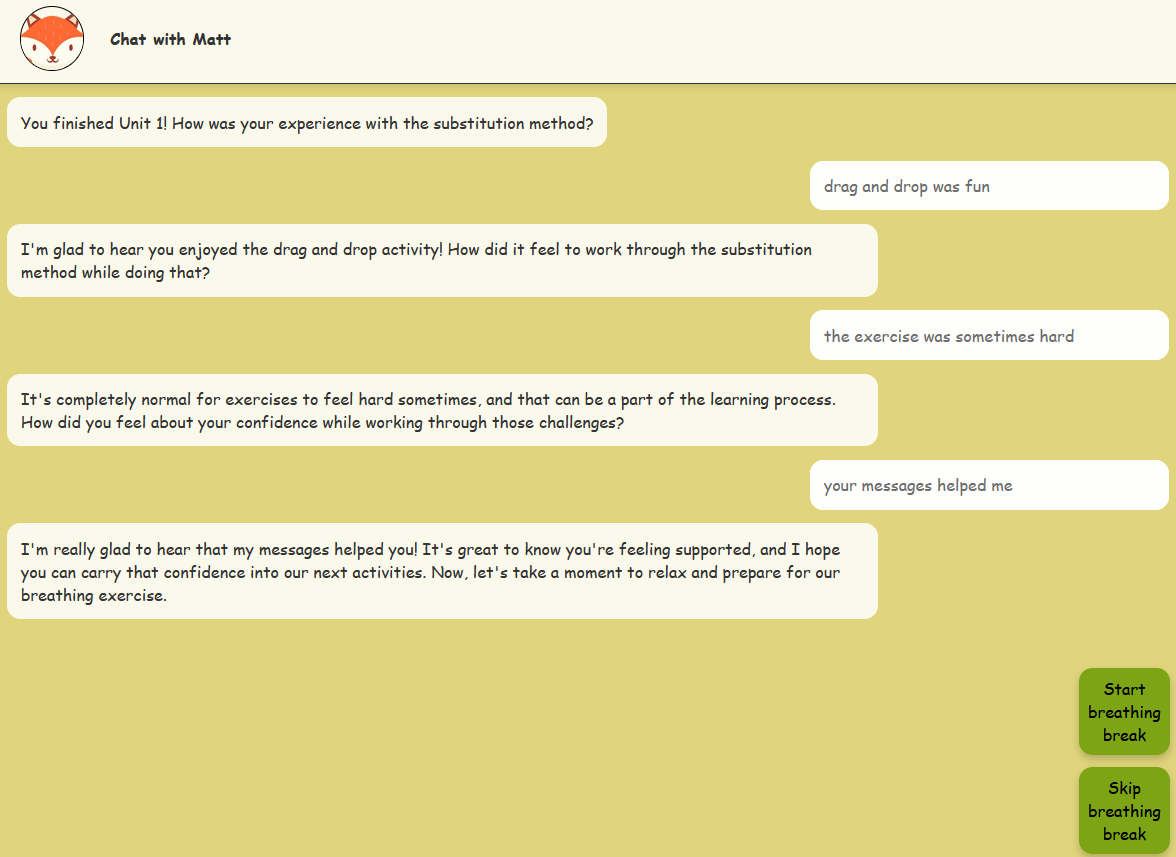}
        \caption{LLM-based chat interaction with skippable breathing exercises}
        \label{fig:chat}
      \end{subfigure}
      \hfill
      \begin{subfigure}[t]{0.49\textwidth}       
        \includegraphics[width=\linewidth]{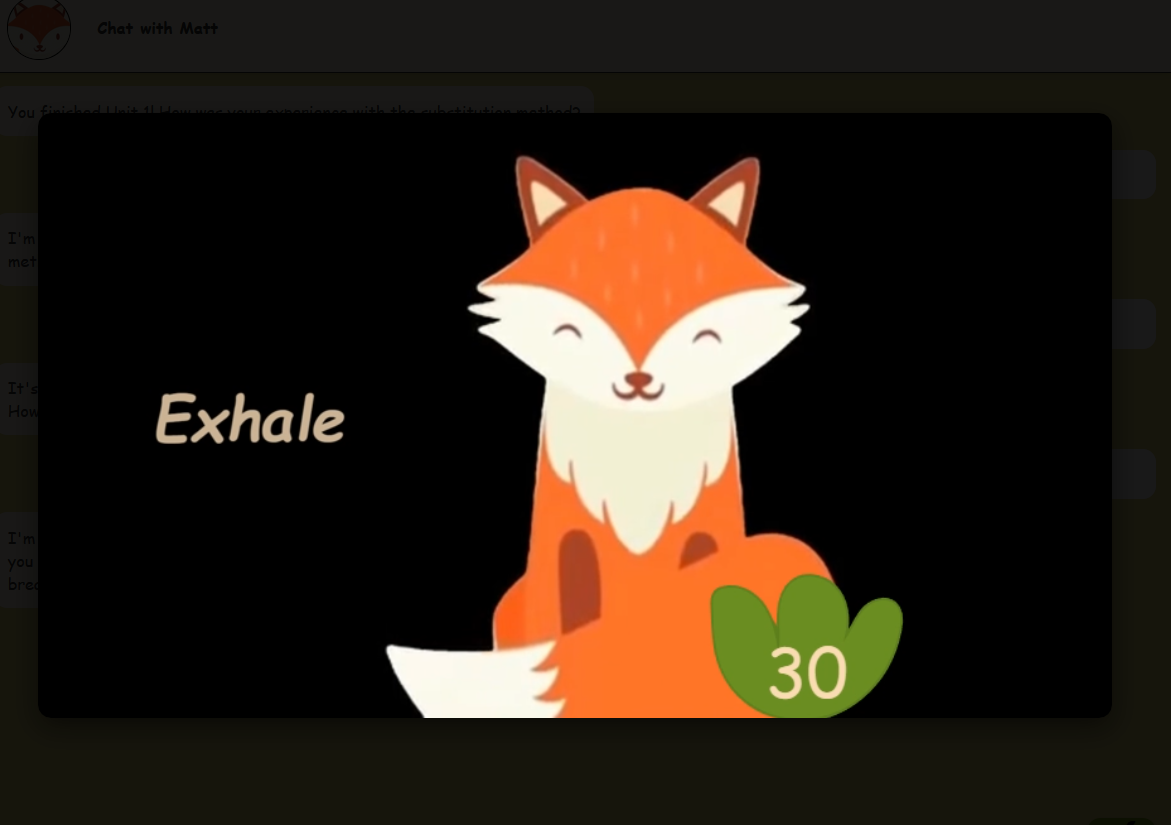}
        \caption{Popup window for breathing exercise}
        \label{fig:breathing}
      \end{subfigure}
  \caption{Math and chat interface}
  \label{fig:mindful}
\end{figure}

\paragraph{Dynamic chats with Matt.}
First, to provide flexible, mindful conversation opportunities to learners, we designed LLM-based flexible chat units, which are introduced between math learning units (Figure \ref{fig:chat}; learners have three math units, and the chats are introduced three times, after each math unit
). In the mindful chat units, students would be able to freely express their internal thoughts and emotions in their own words and have a brief conversation with Matt. This chat interface harnessed OpenAI's GPT-4o-mini model (with a temperature of 0.4 and a maximum of 100 tokens (about two sentences)) via a REST API. Each chat began with a fixed message from Matt, and subsequent messages were generated dynamically, constrained by a limited number of messages, as required by CTAT's interface design and behavior graph logic. 

\begin{figure}
    \centering
    \includegraphics[width=\linewidth]{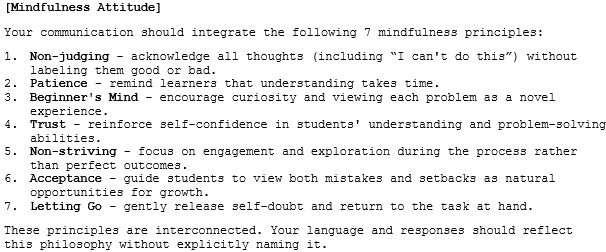}
    \caption{Mindfulness principles in LLM prompt}
    \label{fig:prompt}
\end{figure}

The LLM was prompted to role-play as a warm and supportive math tutor, who applies mindful language in its responses. The system prompt incorporates mindful language based on Kabat-Zinn's seven mindfulness principles \cite{Kabat-Zinn2015}, which translates each principle into specific communicative intentions suitable for an educational context, such as validating a learner’s pace \cite{Hladky2025_MindfulMathTutor}. This prompt guided all chat interactions (Fig. \ref{fig:prompt}) and the generation of hint and feedback messages. Specific prompts for the chat interactions were generated dynamically at runtime to incorporate the current dialog state, chat history and current student’s input. Additionally, it included critical interaction guidelines for handling inappropriate language, nonsensical answers and off-topic responses to ensure appropriate responses for specific cases.

\paragraph{Breathing exercises.}
Following mindfulness intervention studies that show promising effects of breathing exercises \cite{Brunye2013,Hladky2025_MindfulMathTutor,Zenner2014,Zuo2023}, we also implemented a page where students would be guided through a 60-second-long breathing exercise with Matt. Students would be offered to engage in the breathing exercise after completing each mindful chat unit in the tutor (Fig. \ref{fig:chat}, \ref{fig:breathing}).

\paragraph{Mindful hints and feedback messages.}
We refrained from using LLMs in giving dynamic hints and feedback messages during problem solving given potential hallucinations. Yet, we intentionally designed hints and feedback messages to be mindful (not only focusing on cognitive support). Specifically, Matt's messages were first designed based on common ITS feedback
, which we then enhanced with mindful language by prompting an LLM (see \ref{fig:prompt}) to convert cognitively-oriented common ITS feedback into mindful one (Table. \ref{tab:feedback}). We hard-coded these generated messages in the CTAT architecture during the ITS development, after checking for accuracy.
\begin{table}[!h]
    \centering
    \caption{Mindful language enhancement of cognitive hint and feeback message}
    \begin{tabularx}{\textwidth}{l X X} \toprule
         & \textbf{Cognitive Base} & \textbf{Mindful Language Enhancement}  \\
         \midrule
         Success feedback & You've created an equation with just one variable. Let's simplify it next. & Nice work creating an equation with one variable. Let's explore simplification next.  \\
         \addlinespace
         Failure feedback & If you would like my help, click on the hint button. & Not quite there, but that's completely okay! Learning takes time and patience. The hint can help you get moving again. \\
         \addlinespace
         Hint & Combine like terms: 2x+x=3x. & Combining like terms could be the next step.; 2x+x can simplify to 3x. This process shows how we can make equations easier to work with. \\
         \bottomrule
    \end{tabularx}
    \label{tab:feedback}
\end{table} 
\section{Classroom Evaluation}
To empirically evaluate the system, we asked three research questions: 
\begin{itemize}
    \item (\textbf{RQ1}) How do students' state-MA change by interacting with the mindful ITS?
    \item  (\textbf{RQ2}) Do students learn math knowledge and skills using the mindful ITS?
    \item (\textbf{RQ3}) How do students perceive the usability of the mindful ITS, and the pedagogical agent’s effectiveness in providing support, building confidence, and regulating emotions? 
\end{itemize}

We conducted a classroom experiment testing our ITS (Mindful condition) with a version of the same ITS without the mindfulness interventions described above (i.e., LLM-based mindful chat, breathing, and mindful hints and feedback during problem solving), which we called the Cognitive condition. 
Comparing these two systems not only on math anxiety but also math learning presents an intellectually-interesting problem to evaluate whether the mindful chats and breathing (at the cost of math practice time) would influence student learning negatively or not. Note that in both versions, Matt was available throughout the tutor learning; in the Cognitive condition, Matt did not provide any particular mindful support and only provides cognitively-oriented hints and feedback messages (see examples in Table. \ref{tab:feedback}).

\subsection{Participants}
Participants included 252 7th-grade students (aged 12-13) at an international school in Japan across seven classes. Students in each class were randomly assigned to the \textit{Mindful }condition (using the version with the mindfulness interventions) or the \textit{Cognitive} condition (without the interventions). Due to external unexpected disruptions (illness and a major incident on the subway network) that led to the cancellation of class sessions, only two of the seven classes completed all study activities, consisting of 42 students (18 in Mindful, 24 in Cognitive conditions).  Class cancellations occurred at the class level, thus did not affect the distribution of the condition assignment unfairly. The study had been approved by the ethics committee of an affiliated university prior to data collection. 

\subsection{Measures}
\paragraph{Math learning.}
To measure students' math learning, we conducted pretest and posttest on students' conceptual and procedural knowledge about linear systems of equations. Each test contained six items.

\paragraph{Trait math anxiety (Trait-MA).}
Trait-MA was assessed by adopting the Abbreviated Math Anxiety Scale (AMAS) \cite{Hopko2003_AMAS}. Participants rated their trait-MA on eight items (e.g., \textit{"Taking an examination in a math course"}) on a 5-point scale (where 5= "very anxious"). We used this data to sort students into low, medium and high anxiety groups for later analyses.

\paragraph{State math anxiety (State-MA).}
State-MA was evaluated using the six-item version of the state scale of the State-Trait Anxiety Inventory (STAI) \cite{Marteau1992_STAI}. Participants rated their current emotional state (e.g. \textit{"I am worried"}) on a 5-point Likert scale (1=not at all, 5=very much). We administered it before math pre/posttest to evaluate anticipatory state-MA, as well as during math pre/posttest (after 4 out of 6 items) to assess executive state-MA.

\paragraph{System usability.}
We adopted the System Usability Scale (SUS) \cite{Brooke1995_SUS}. Participants rated statements (e.g., \textit{"I thought the system was easy to use"}) on a 5-point Likert scale (1=not at all, 5=very much).

\paragraph{Evaluation of the pedagogical agent.}
The questionnaires for evaluating the pedagogical agent included four items each, rated on a 5-point Likert scale (1=not at all, 5=very much) for perceived socio‑emotional support (e.g., \textit{"I felt comfortable learning with Matt"}) \cite{Hladky2025_MindfulMathTutor}, warmth (e.g., \textit{"I believe that Matt acts in my best interest"}) and competence (e.g, \textit{"I believe Matt is very capable of performing his job"}). Items for warmth and competence were adapted from the Human Computer Trust (HCT) \cite{Madsen2000_Competence}, Trust \cite{Merritt2011_Competence}, and related measures \cite{Calhoun2019_Competence_Warmth,Li2025_Competence_Warmth}. 

\paragraph{Log data.} To gain additional exploratory insights, we also logged various kinds of student-ITS interactions (e.g. button clicks, hint requests, error rates and and chat messages).

\subsection{Procedure}
The study was embedded in a weekly 50-minute "Information Technology and Data Science" class at the school, taught by one teacher. The study consisted of two sessions across two weeks. Session 1 included a 10 min introductory presentation, pretest (15 min, Tab. \ref{tab:questionnairs}) and 15 minutes of ITS use. Session 2 started with 15 minutes of continued ITS use, followed by posttest (15 min, Tab. \ref{tab:questionnairs}) and a short debriefing presentation at the end of the study. The study was conducted in English, with necessary translation support provided in Japanese when students asked for help.

\begin{table}[htpb]
    \centering
    \caption{Overview of the pre- and posttest questionnaires}
    \begin{tabularx}{\textwidth}{X X} \toprule
        \textbf{Pretest: Start of session 1} & \textbf{Posttest: End of session 2}\\
        \midrule
         Demographics (Age, gender, English level)& \\
         Trait-MA & Trait-MA \\
         Anticipatory MA & Anticipatory MA\\
         Math performance incl. executive MA & Math performance incl. executive MA \\
         & System usability (SUS)\\
         & Matt's characteristics\\
         \bottomrule
    \end{tabularx}
    \label{tab:questionnairs}
\end{table}

\section{Results}
This section reports the findings for the three research questions regarding trait- (Abbreviated Math Anxiety Scale (AMAS) \cite{Hopko2003_AMAS}) and state-MA (State-Trait Anxiety Inventory (STAI) \cite{Marteau1992_STAI}) (RQ1), algebra learning (math test with six items) (RQ2), and learner perceptions of the system (System Usability Scale (SUS) \cite{Brooke1995_SUS}) and the pedagogical agent (RQ3). To gain additional exploratory insights, we also logged various kinds of student-ITS interactions (e.g. button clicks, hint requests, error rates and and chat messages). All tests satisfied assumption checks (normality, homoscedasticity, and outlier diagnostics) unless otherwise noted. 
Though the final sample who completed all study activities was small (N=42)  we complement it by providing additional insights from tutor log data, which were collected from all 252 participating students.


\subsection{Math Anxiety}
\paragraph{Trait-MA.} The AMAS yielded a broad score range from 8 to 31 (\textit{M}=18.3, \textit{SD}=6.85, possible range: 8 to 40). A Mann-Withney-U test (due to non-normality in the Cognitive group) showed no statistically significant group difference (\textit{W}=224, \textit{p}=.849).
Students were categorized into three anxiety levels based on their AMAS score using established cutoff thresholds~\cite{Ashcraft2001_Anx_and_Perform}. The majority of students (\textit{N}=27) were categorized as \textit{medium anxiety}, \textit{N}=9 as \textit{low anxiety} and \textit{N}=6 as \textit{high anxiety}. 

\paragraph{State-MA.} Anticipatory and executive state‑MA were measured by the STAI‑6 (possible range: 20 to 80). We administered it before math pre/posttest to evaluate anticipatory state-MA, as well as during math pre/posttest (after 4 out of 6 items) to assess executive state-MA. Table~\ref{tab:state-ma} shows descriptive state-MA measures for both Cognitive and Mindful condition before and after the intervention (pre-/post-test). Both groups show comparable values in the pre-test. The Mindful group shows a decrease in executive MA but an increase in anticipatory MA, while the Cognitive group shows a decrease in both MA types. A linear mixed‑effect model revealed no statistically significant effects of group, time, and interaction between groups and time for anticipatory state‑MA. However for executive state‑MA, both groups showed a significant main effect for time with a large effect size (\textit{F}(1, 15)=5.422, \textit{p}=.034, partial $\eta^2$=0.27), but no group nor interaction effect was found. Overall state‑MA (sum of anticipatory and executive scores) changed neither by time nor by group. Thus, the Mindful ITS version did not significantly change the anticipatory or executive state-MA when compared against the Cognitive ITS version without the mindfulness interventions. 

\begin{table}[htpb]
    \caption{State-MA mean and standard deviation by group and time (possible range: 20 to 80)}
    \label{tab:state-ma}
    \begin{tabularx}{\textwidth}{X X X X} \toprule
    \textbf{Group} & \textbf{State-MA Type} & \textbf{Pretest} (\textit{M(SD)}) & \textbf{Post-test} (\textit{M(SD)})\\
    \midrule
    Cognitive & Anticipatory & 41.67 (13.48) & 37.64 (13.78) \\
              & Executive & 43.64 (14.79) & 33.64 (9.48) \\
    \addlinespace
    Mindful   & Anticipatory & 40.37 (13.57) & 44.63 (16.61) \\
              & Executive & 38.89 (13.93) & 33.89 (14.82) \\
    \bottomrule
    \end{tabularx}
\end{table}

\subsection{Math Learning}
The pre/posttests each consisted of six questions and students could score maximum nine points. The groups did not differ on pretest (\textit{t}(39.48)=-0.024, \textit{p}=.981, \textit{d}=-0.007) and showed improvements from pretest to posttest: the Cognitive condition increased from \textit{M}=4.38 (\textit{SD}=2.06) to \textit{M}=5.33 (\textit{SD}=2.08) and the Mindful condition from \textit{M}=4.39 (\textit{SD}=1.72) to \textit{M}=4.72 (\textit{SD}=1.93). A linear mixed-effect model revealed a significant main effect of time with a large effect size (\textit{F}(1, 40)=6.418, \textit{p}=.015, partial $\eta^2$=.14), but no group or interaction effect, indicating that students learned algebra knowledge, but the improvement did not depend on the condition.

\subsection{System Usability}
System usability was rated relatively low in both groups (possible range: 0 to 100). The Cognitive group's rating of \textit{M}=43.96 (\textit{SD}=26.06) corresponded to an F grade according to the curved grading scale by Lewis et al. \cite{Lewis2018_SUS_details}, while the Mindful group's rating of  \textit{M}=55.56 (\textit{SD}=21.05) corresponded to a D grade. Although the Mindful version received higher SUS ratings than the Cognitive version, the difference was not statistically significant (\textit{t}(39.74)=-1.594, \textit{p}=.119, \textit{d}=-0.48).

\subsection{Pedagogical Agent's Characteristics}
The questionnaires for evaluating the pedagogical agent included four items each, rated on a 5-point Likert scale (1=not at all, 5=very much) for perceived socio‑emotional support (e.g., \textit{"I felt comfortable learning with Matt"}) \cite{Hladky2025_MindfulMathTutor}, warmth (e.g., \textit{"I believe that Matt acts in my best interest"}) and competence (e.g, \textit{"I believe Matt is very capable of performing his job"}). Items for warmth and competence were adapted from the Human Computer Trust (HCT) \cite{Madsen2000_Competence}, Trust \cite{Merritt2011_Competence}, and related measures \cite{Calhoun2019_Competence_Warmth,Li2025_Competence_Warmth}. 
Table \ref{tab:character} summarizes the questionnaire scores for both groups (possible range for each: 4 to 20). Since these questionnaires were self-composed for this study, Crohnbach's alpha was computed for each of them. They all showed a high value (socio-emotional support $\alpha$: 0.88, warmth: $\alpha$ = 0.94, competence: $\alpha$= 0.91), suggesting that the items within each scale are highly inter-related and measuring the same underlying construct.

The Mindful group reported higher scores on all three dimensions, but overall they were not significantly different from those of the Cognitive group. 
To gain deeper insights, we performed single‑item analyses on each questionnaire, revealing that the item "I feel supported by Matt" (socio-emotional support questionnaire) was significantly higher rated in the Mindful condition (\textit{W}=167.0, \textit{p}=.028), as well as the item “I believe that Matt is very concerned about my well-being” (warmth questionnaire) (\textit{W}=139.5, \textit{p}=.046).
\begin{table}[htpb]
    \caption{Perceived socio-emotional support, warmth and competence mean and standard deviation of pedagogical agent Matt by group (possible range: 4 to 20)}
    \label{tab:character}
    \begin{tabularx}{\textwidth}{X X X} \toprule
        \textbf{Characteristics} &  \textbf{Mindful} (\textit{M(SD)}) & \textbf{Cognitive} (\textit{M(SD)})\\
        \midrule
       Socio-Emotional Support & 11.72 (4.61) & 9.25 (3.79) \\
       Warmth & 12.06 (5.35) & 9.88 (4.60) \\
       Competence & 12.61 (5.32) & 10.96 (4.51) \\
       \bottomrule
    \end{tabularx}
    
\end{table}

\subsection{Additional Insights: Exploratory Log Data Analysis}
We collected ITS log data to compare differences between the two groups in ITS-interaction behavior. Table \ref{tab:log} shows descriptive statistics on students' averaged problem-solving steps completed,, their hint request rate, and error rate. 
\begin{table}[htpb]
    \caption{Learning process data from the tutor interactions}
    \label{tab:log}
    \begin{tabularx}{\textwidth}{l X X X} \toprule
        \textbf{Group} & \textbf{Total Steps} (\textit{M(SD)}) & \textbf{Hint Rate} (\textit{M(SD)}) & \textbf{Error Rate} (\textit{M(SD)})\\
        \midrule
        Cognitive & 21.28 (11.61) & 1.06 (1.17) & 0.46 (0.38) \\
        \addlinespace
        Mindful & 12.43 (8.88) & 0.61 (0.76) & 0.47 (0.45)\\
        \bottomrule
    \end{tabularx}
\end{table}

\paragraph{Total number of problem-solving steps.}
The intervention consisted of three math units (57 steps) and ten bonus problems (65 steps) with a total of 122 problem-solving steps.
Students in the Cognitive group completed on average 21.28 math solving steps, which corresponded to the beginning of the second problem in Unit 2. Students in the Mindful group completed on average 12.43 math solving steps, which corresponded to the middle of the third problem in Unit 1. The difference between the groups was statistically significant (Mann-Whitney-U, \textit{W} = 11930, \textit{p} < .001). 

\paragraph{Hints per completed problem-solving step.}
Each step had three hints available, of which students in the Cognitive group used \textit{M} = 1.06 hints per step and students in the Mindful group used \textit{M} = 0.61 hints. This group difference was significant (Mann-Whitney-U, \textit{W} = 8645, \textit{p} = .002).

\paragraph{Errors per completed problem-solving step.}
Error rate did not significantly differ between groups with \textit{M} = 0.46 (\textit{SD} = 0.38) for the Cognitive and \textit{M} = 0.47 (\textit{SD} = 0.45) for the Mindful group (Mann-Whitney-U, \textit{W} = 7257, \textit{p} = .606). 
 
\section{Discussion}
This study explored the integration of LLM-based mindfulness interventions into a math Intelligent Tutoring System (ITS). By embedding a pedagogical agent (Matt) capable of delivering context-sensitive emotional support through mindful chat interactions and breathing exercises, we investigated whether such socio-emotional scaffolding could enhance traditional cognitive tutoring and reduce math-related negative emotions.
The majority of our student sample reported to generally experience anxiety when faced with mathematics, with one in seven reporting high math anxiety (MA), reflecting past PISA study reports~\cite{OECD2015} and indicate a need for action to implement emotion regulation supporting measures into regular math practice.

Our findings demonstrate that while the addition of mindfulness components did not yield significantly different outcomes in math anxiety or learning compared to cognitive support alone, the system successfully established a perceived socio-emotional connection with learners. Further, tutor log data revealed that students with the mindfulness interventions worked on significantly fewer steps 
compared to their peers without the interventions. It is notable, however, that students with the mindfulness interventions demonstrated learning outcomes that are no different from students who were only working on math practice in the ITS. In the following, we discuss our findings in detail.

We investigated how students' momentary experience of anxiety in a math-related context (state-MA) changes by interacting with a mindful ITS (RQ1). For experienced MA in anticipation of a math interaction (anticipatory state-MA), we did not find a significant reduction over time or difference between groups. However, executive state-MA (experienced during a math interaction) decreased significantly over time for both groups. This lack of between-group difference may be caused by constrains in the classroom environment. While ecologically valid, the social dynamics can undermine individual mindfulness engagement. Students may have felt uncomfortable about closing their eyes for breathing exercises while surrounded by peers, and we observed distracting conversations between students comparing interface versions. This highlights a critical challenge for digital mindfulness interventions in public classroom settings. Our self-paced digital approach was vulnerable to peer influence, potentially diluting the intervention's affective impact.

Furthermore, we analyzed how mindfulness interventions affect students' learning outcomes (RQ2). Both groups significantly improved their math performance from pretest to posttest with no condition differences, indicating that the shared ITS elements, like instructions, hints and feedback messages, and overall design of the math exercises were effective in increasing math performance, consistent with established work on ITSs \cite{Koedinger2007,Lehn2011}. 
Additional exploratory analyses on students' learning processes using system log data revealed that students that experienced the mindfulness interventions solved significantly fewer problems overall and used fewer hints per solving step. This indicates benefits for learning efficiency and self-regulation: Even with fewer opportunities for practicing math problems (they spent a decent amount of time chatting and breathing with Matt), they achieved similar math learning outcomes like students that used an ITS version focused on math learning only while using a well-balanced amount of help needed to achieve meaningful learning.
Note that during chatting and breathing exercises, there was no math-related support provided (e.g., Matt would not give any cognitive support during the LLM-supported chat).

Finally, we investigated the system's usability and how students perceived the pedagogical agent (RQ3). System usability was rated as low, likely caused by students using browser-integrated translation tools due to limited English proficiency. Observation showed that using such tools lead to some interface elements becoming unresponsive. 
Conversely, the pedagogical agent Matt was very well received independently of the group. Still, evaluation of individual questionnaire items revealed that students that experienced the mindful interventions perceived Matt as more supportive, empathetic and caring, indicating socio-emotional benefits of the mindful chat interactions.

\subsection{Limitations and Future Work}
We acknowledge several limitations. First, participants lacked sufficient English proficiency to fully understand interface elements (e.g., buttons), as well as hints, feedback, and chat messages. Students managed to complete tasks by using translation tools, but this disrupted students' workflow and may have affected their experience.
This limitation likely masked potential effects of the intervention -- if students cannot fully comprehend the mindful language embedded in hints and chat, the affective scaffolding cannot function as intended. In an ongoing re-implementation we prioritize full linguistic and cultural localization to realize the benefits of LLM-based socio-emotional support.

Second, the sample size for evaluating changes in math anxiety and math performance was smaller than expected due to external constraints and therefore the results may not generalize across contexts.
Further, the classroom setting appeared to limit the engagement in individual, self-paced mindfulness practices (e.g., students would make fun of others). Future studies could mitigate this by either using randomized group assignment at the class level to reduce peer pressure while maintaining the classroom context, or by integrating the study activities as homework assignments. 

\subsection{Conclusion}
The present work integrated mindful LLM-based chat interactions, breathing exercises, and mindful hints and feedback in an ITS through a pedagogical agent. We conducted a classroom experiment to test the effect of these interventions on students' self-reported anxiety, math learning, and their perception towards the pedagogical agent.
The findings indicate that the overall learning experience with the ITS contributed to a reduction in executive state-math anxiety and an increase in math learning. While the classroom implementation and language barrier faced challenges that obscured differential effects, the findings reveal promising directions. Mindfulness interventions may promote learning efficiency and self-regulated problem-solving behaviors, while LLM-based pedagogical agents can successfully establish perceived emotional support and warmth. 

This study demonstrates the feasibility of integrating LLM-driven mindfulness into structured Intelligent Tutoring Systems, offering a scalable approach to socio-emotional support that does not require specialized instructor training. While the results underscore the potential of such approaches, they also highlight the need for further research to better understand how mindfulness components can be optimally embedded into ITS and educational context to achieve measurable and robust effects on math anxiety and performance.  

%
%
%
\bibliographystyle{splncs04}
\bibliography{reference}

\end{document}